\documentclass{kapproc} 

\usepackage{t1enc}

\usepackage{procps} 

\usepackage[dvips]{graphicx}

\upperandlowercase

\let\footnote\savefootnote

\let\footnotetext\savefootnotetext

\let\footnoterule\savefootnoterule

\kluwerbib 

\def\kms{\relax \ifmmode {\,\rm km\,s}^{-1}\else \,km\,s$^{-1}$\fi}
\def\ha{\relax \ifmmode {\rm H}\alpha\else H$\alpha$\fi}
\def\hb{\relax \ifmmode {\rm H}\beta\else H$\beta$\fi}
\def\hi{\relax \ifmmode {\rm H\,{\sc i}}\else H\,{\sc i}\fi}
\def\hii{\relax \ifmmode {\rm H\,{\sc ii}}\else H\,{\sc ii}\fi}
\def\h2{\relax \ifmmode {\rm H}_2\else H$_2$\fi}

\def\fdg{\hbox{$.\!\!^\circ$}}

\def\farcm{\hbox{$.\mkern-4mu^\prime$}}
\def\farcs{\hbox{$.\!\!^{\prime\prime}$}}
\def\degd#1.#2{ #1\fdg#2 }                 
\def\mind#1.#2{ #1\farcm#2 }               
\def\secd#1.#2{ #1\farcs#2 }               

\def\aj{AJ}                   
\def\araa{ARA\&A}             
\def\apj{ApJ}                 
\def\apjl{ApJ}                
\def\apjs{ApJS}               
\def\aap{A\&A}                
\def\mnras{MNRAS}             
\def\nat{Nature}              

\begin{document}

\articletitle[Fuelling starbursts and AGN]
{Fuelling starbursts and AGN}

\author{Johan H. Knapen}

\affil{Centre for Astrophysics Research, University of Hertfordshire,
Hatfield, Herts AL10 9AB, UK}

\begin{abstract}

There is considerable evidence that the circumnuclear regions of
galaxies are intimately related to their host galaxies, most directly
through their bars. There is also convincing evidence for relations
between the properties of supermassive black holes in the nuclei of
galaxies and those of their host galaxy. It is much less clear,
however, how stellar (starburst) and non-stellar (AGN) activity in the
nuclear regions can be initiated and fuelled. I review gas transport
from the disk to the nuclear and circumnuclear regions of galaxies, as
well as the statistical relationships between the occurrence of
nuclear activity and mechanisms which can cause central gas
concentration, in particular bars and interactions. There are strong
indications from theory and modelling for bar-induced central gas
concentration, accompanied by limited observational evidence. Bars are
related to activity, but this is only a weak statistical effect in the
case of Seyferts, whereas the relation is limited to specific cases in
starbursts. There is no observational evidence for a statistical
connection between interactions and activity in Seyferts, and some
evidence for this in starbursts, but probably limited to the extremes,
e.g., ULIRGs. Some interesting hints at relations between rings,
including nuclear rings, and the presence of nuclear activity are
emerging.  It is likely that the connection between the inflow of
gaseous fuel from the disk of a galaxy on the one hand and the
activity in its nuclear region on the other is not as straightforward
as sometimes suggested, because the spatial- or time-scales concerned
may be significantly different.

\end{abstract}


\section{Introduction}

Black holes are ubiquitous in the nuclei of both active and non-active
galaxies (e.g., Kormendy \& Richstone 1995), and are thought to be at
the direct origin of non-stellar nuclear activity (e.g., Lynden-Bell
1969; Begelman, Blandford \& Rees 1984). The fact that the mass of the
central supermassive black hole (SMBH) in a galaxy is correlated with
the velocity dispersion of the bulge, and hence with its mass
(Ferrarese \& Merritt 2000; Gebhardt et al. 2000) provides the most
tangible link between the nuclear regions and their host galaxies. But
because not all galaxies with SMBHs have AGN characteristics, the
presence of an SMBH in itself cannot be enough to make a galaxy
``active'', at least not continuously, and additional mechanisms must
be considered which can ignite the nuclear activity.

In the case of starburst galaxies, defined rather loosely as galaxies
which show abnormally enhanced massive star formation activity in
their central regions (or in some more extreme cases throughout the
galaxy), a similar question can be posed, namely what ignites the
starburst. In both the AGN and the starbursts, the availability of
fuel at the right place and at the right time must play a critical
role. Such gaseous fuel is plentiful in the disks of galaxies, but
must lose significant quantities of angular momentum in order to move
radially inward. In fact, the ``fuelling problem'' is not the amount
of fuel that is available, but how to get it to the right place, as
graphically illustrated by Phinney (1994, his fig.~1). Estimates for
the mass accretion rate needed to fuel AGN vary from around
$10^{-4}$~M$_\odot$/year for low-luminosity AGN such as LINERs, up to
around 10~M$_\odot$/year for high-luminosity AGN such as QSOs, or,
over a putative lifetime of 10$^8$~years for the AGN activity, only
10$^4$ to $10^9$~M$_\odot$.

Large stellar bars, as well as tidal interactions and mergers, have
some time ago been identified as prime candidates to drive gas
efficiently from the disk into the inner kpc (see next Section). In
this review we will concentrate on the observational evidence, mostly
statistical in nature, for the effectiveness of these gravitational
mechanisms, concentrating on the effects of bars in Section~2, and on
those of interactions in Section~3. Galactic rings are considered in
Sections 4 and 5, and summarising remarks are given in
Section~6. Related reviews considering the fuelling of primarily AGN
include those by Shlosman, Begelman \& Frank (1990), Beckman (2001),
Combes (2001), Shlosman (2003), Wada (2004), and Jogee (2004).

\section{The effects of bars}

\begin{figure}[ht]
\vskip.2in
\centerline{\includegraphics[width=4in]{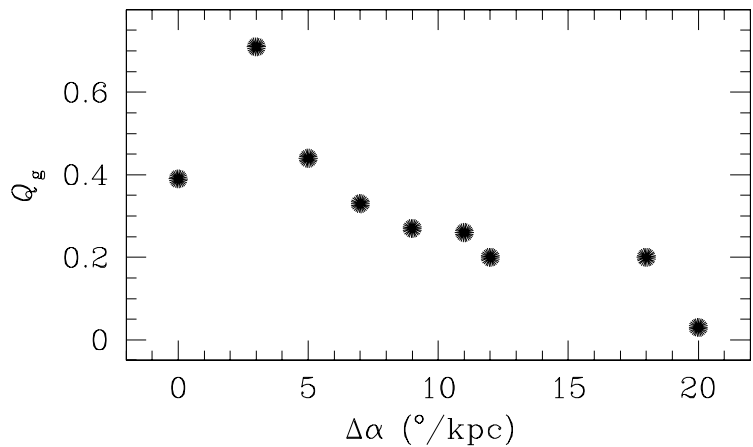}}
\caption{Gravitational bar torque $Q_{\rm g}$, an indicator of bar
strength, as a function of the curvature of the dust lanes
$\Delta\alpha$ in a sample of 9 barred galaxies. Small values of
$\Delta\alpha$ indicate straight dust lanes, which are seen to occur
in strong bars, thus confirming theoretical and numerical
predictions. Data from Knapen, P{\' e}rez-Ram{\'{\i}}rez \& Laine
(2002).}
\label{dustlanes}
\end{figure}

Theoretically and numerically, bars are expected to concentrate gas in
the central regions of spiral galaxies because the torqued and shocked
gas within the bar loses angular momentum which allows the gas to move
further in (e.g., Schwarz 1984; Combes \& Gerin 1985; Noguchi 1988;
Shlosman, Frank \& Begelman 1989; Knapen et al. 1995a). The dynamics
of bars and their influence on the circumnuclear regions has most
recently been reviewed by Kor\-men\-dy \& Kennicutt (2004), and
previously by, e.g., Sellwood \& Wilkinson (1993) and Shlosman
(2001). The general theoretical and numerical formalism of bars is now
well understood, and different aspects of it are continuously being
confirmed by observations. For instance, we recently investigated the
well-known numerical result that stronger bars will lead to straight
dust lanes along the leading edges of the bar, whereas the dust lanes
will be more curved in weak bars (Athanassoula 1992). Using a small
number of barred galaxies for which we had adequate data, we could
indeed confirm observationally that there is an anti-correlation
between the amount of curvature of the dust lanes and the
gravitational bar torque, or bar strength (Knapen, P\'erez-Ram\'\i rez
\& Laine 2002; see Fig.~\ref{dustlanes}). In another study (Zurita et
al. 2004), we used \ha\ Fabry-P\'erot data of the strongly barred
galaxy NGC~1530 to show in a graphic, two-dimensional way that indeed,
as predicted by theory, large velocity gradients are found at the
position of the dust lanes. Within those lanes, directly tracing
enhanced concentrations of dust and thus gas, but indirectly tracing
the location of shocks in the gas, the large velocity gradient
prohibits massive star formation, which we observe to be located just
outside the regions of largest shear or velocity gradient (Zurita et
al. 2004; see Regan, Vogel \& Teuben 1997 for an \ha\ Fabry-P\'erot
map at lower resolution which nevertheless indicates the shocks in the
velocity field).

Considering now specifically the theoretical and numerical view that
bars can instigate radial inflow of gas, and thus lead to gas
accumulation in the central regions of barred galaxies, several pieces
of observational evidence to fit this picture have been forthcoming in
recent years, both from observations of gas tracers in barred and
non-barred galaxies (e.g., Sakamoto et al. 1999; Jogee, Scoville, \&
Kenney 2004; Sheth et al. 2004), and from other, less direct, measures
of the gas concentration (e.g., Maiolino, Risaliti, \& Salvati 1999;
Alonso-Herrero \& Knapen 2001). These results have been reviewed in
somewhat more detail by Knapen (2004a), and we limit ourselves here to
the conclusion that there is increasing observational support for the
theoretical suggestion that bars lead to gas accumulation in the
central regions of galaxies.

One must keep in mind that in all cases the observed correlation
between the presence of a bar and increased central gas concentration
is {\it statistical}, and not very strong, that there is a large
overlap region in which the properties of barred and non-barred
galaxies are very similar indeed (for instance, in the study by
Sakamoto et al. 1999 just over half of the 19 sample galaxies are in
the overlap range of gas concentration parameter $t_{\rm con}$, which
is inhabited by both barred and non-barred galaxies), and that in the
CO studies the $X$ factor which gives the transformation of CO
luminosity to mass is assumed to have the same value in the
circumnuclear regions and in the disk. One can also question whether
the statistical gas accumulation by bars is in fact related to the
occurrence of nuclear activity of the non-stellar or stellar variety,
and a careful consideration of both spatial- and timescales must be
made to connect gravitationally driven inflow to fuelling of the
starburst and/or the AGN. Finally, there is as yet no convincing
direct observational evidence of inflow in a barred galaxy, mainly
because the inflow rates are so low that they may be unobservable in
practive (see above), and because most of the gas in bars moves around
the bar, and will thus move inward during a part of its orbit, but
then move outward again on a subsequent part (see discussion in Knapen
2001).

In the remainder of this Section, we will review the question of
whether there is observational evidence that bars are related to the
occurrence of nuclear activity.

\subsection{Bars and starburst activity}

There is a clearly observed trend for nuclear starbursts to occur
preferentially in barred hosts (e.g., Hummel 1981; Hawarden et
al. 1986; Devereux 1987; Dressel 1988; Puxley, Hawarden, \& Mountain
1988; Arsenault 1989; Huang et al. 1996; Martinet \& Friedli 1997;
Hunt \& Malkan 1999; Roussel et al. 2001).  For example, Hummel (1981)
reported that the central radio continuum component is typically twice
as strong in barred as in non-barred galaxies; Hawarden et al. (1986)
found that barred galaxies dominate the group of galaxies with a high
25$\mu$m/12$\mu$m flux ratio; and Arsenault (1989) found an enhanced
bar$+$ring fraction among starburst hosts. Huang et al. (1996)
revisited IRAS data to confirm that starburst hosts are preferentially
barred, but did point out that this result only holds for strong bars
(SB class in the RC3 catalogue, de Vaucouleurs et al. 1991) and in
early-type galaxies, results confirmed more recently by Roussel et
al. (2001). In contrast, Isobe \& Feigelson (1992) did not find an
enhanced far-IR to blue flux ratio among barred galaxies, and Ho,
Filippenko \& Sargent (1997) found only a very marginal increase in
the detection rate of \hii\ nuclei (indicative of starburst activity)
among the barred as compared to non-barred galaxies in their sample,
only among the late-type spirals (Sc-Sm), and most likely 
resulting from  selection effects rather than bar-induced inflow (Ho et
al. 1997). All results mentioned above rely on optical catalogues such
as the RC3 to derive the morphological classifications, whereas it is
now well-known that the presence of a bar can be deduced more reliably 
from near-IR imaging (e.g., Scoville et al. 1988; Knapen et
al. 1995b). Although near-IR imaging leads to enhanced bar fractions
as compared to optical imaging (e.g., Knapen, Shlosman \& Peletier
2000; Eskridge et al. 2000), it is not clear how it would affect the
results on bars and starbursts.

The statistical studies referred to above thus seem to show that bars
and starbursts are connected, but that the results are subject to
important caveats and exclusions. Further study is needed, determining
bar parameters from near-IR imaging, using carefully defined samples,
and exploring more direct starburst indicators than the IRAS fluxes
which have often been used. Higher-resolution imaging of the starburst
galaxies is also needed, to confirm the possible {\it circum}nuclear
nature of the starburst, already suggested back in 1986 by Hawarden et
al.

\subsection{Bars and Seyfert activity}

\begin{figure}[ht]
\vskip.2in
\centerline{\includegraphics[width=4in]{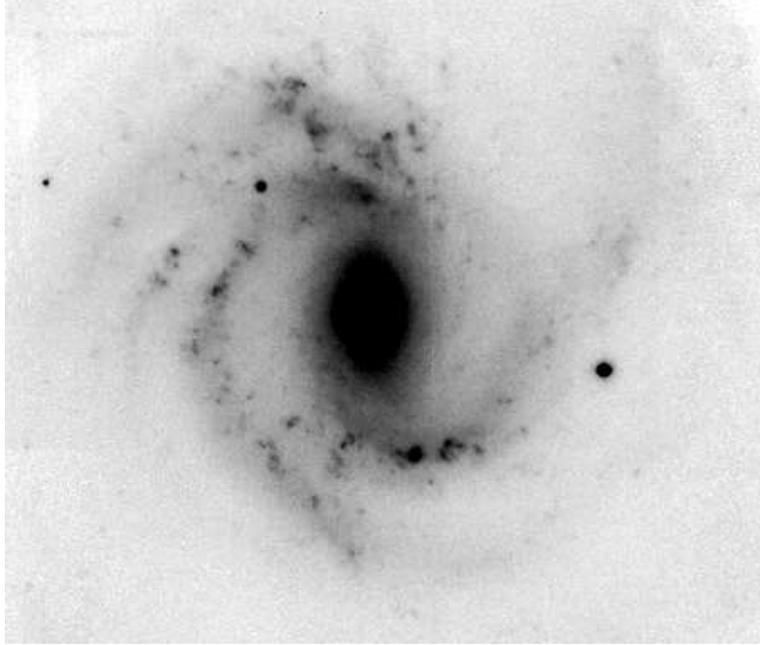}}
\caption{Near-IR $K_{\rm s}$ image of the Seyfert host galaxy
NGC~4303, showing a prominent bar and a well-structured set of spiral
arms. North is up, East to the left, and the size of the image shown
is approximately four minutes of arc. The image was taken with the
INGRID camera on the William Herschel Telescope, see Knapen et
al. (2003) for more details.}
\end{figure}

Seyferts are almost ideal for a study of AGN host galaxies: they are
relatively local and occur predominantly in disk galaxies (see Fig.~2
for a nice example). One of the aspects of the host galaxy -- Seyfert
activity connection that has received a good deal of attention over
the years is the question of whether Seyfert hosts are more often
barred than non-Seyferts.  Starting with the work of Adams (1977),
many authors have dedicated efforts to resolve this question, without
finding conclusive evidence (e.g., Adams 1977; Simkin, Su, \& Schwarz
1980; Balick \& Heckman 1982; MacKenty 1990; Moles, M\'arquez, \&
P\'erez 1995; Ho et al. 1997; Crenshaw, Kraemer, \& Gabel
2003). Unfortunately, many of these investigations are plagued by the
absence of a properly matched control sample, by the use of the RC3
classification or, worse perhaps, ad-hoc and non-reproducible
classification criteria to determine whether a galaxy is barred; and
all of them are based on optical imaging.

Near-IR imaging is much better suited for finding bars (see
Section~2.1), and a small number of studies have combined the use of
high-quality, near-IR imaging with careful selection and matching of
samples of Seyfert and quiescent galaxies. One such study, by Mulchaey
\& Regan (1997), reports identical bar fractions, but Knapen,
Shlosman, \& Peletier (2000), using imaging at higher spatial
resolution and a rigorously applied set of bar criteria, find a
marginally significant difference, with a higher bar fraction in a
sample of CfA Seyferts than in a control sample of non-Seyferts
(approximately 80\% vs. 60\%). The results found by Mulchaey \& Regan
(1997) can be reconciled with those reported by Knapen et al. (2000)
by considering the lower spatial resolution employed by the former
authors.

Laine et al. (2002) later confirmed this difference at a 2.5 $\sigma$
level by increasing the sample size and using high-resolution {\it
HST} NICMOS near-IR images of the central regions of all their active
and non-active sample galaxies. Laine et al. (2002) also found that
almost one of every five sample galaxies, and almost one of every
three barred galaxies, has more than one bar. The nuclear bar
fraction, however, is not enhanced in Seyfert galaxies as compared to
non-Seyferts (see also Erwin \& Sparke 2002).

In a recent paper, Laurikainen, Salo \& Buta (2004) study the bar
properties of some 150 galaxies from the Ohio State University Bright
Galaxy Survey (Eskridge et al. 2002) in terms of their nuclear
properties, among other factors. Using several bar classification
methods on optical and near-IR images, they find that only a Fourier
method applied to near-IR images leads to a significant excess of bars
among Seyferts/LINERs as compared to non-active galaxies: a bar
fraction of 71\%$\pm$4\% for the active/starburst galaxies, versus
55\%$\pm$5\% for the non-active galaxies (at a significance level of
2.5$\sigma$)\footnote{It is an interesting exercise to add the numbers
found by Laurikainen et al. (2004) to those found by Laine et
al. (2002), which would give largely the same overall result in terms
of bar fractions, but with smaller error bars thanks to the increased
sample sizes, and an overall significance level of more than
3$\sigma$. Formally this is not allowed though because the original
samples have been selected using different methods, and should not
simply be added.}.

The Fourier method employed by Laurikainen et al. (2004) is objective
but sensitive largely to classical bars with high surface
brightnesses, and in this respect similar to the strict criteria
applied by Knapen et al. (2000) and Laine et al. (2002), who basically
rely upon significant a rise and fall in a radial ellipticity profile
for bar identification. The results of the Fourier analysis by
Laurikainen et al. (2004) are remarkably similar to those from Knapen
et al. and Laine et al., apparently because all trace prominent bars
in near-IR images. Laurikainen et al. find that the excess of bars
among Seyferts/LINERs does not manifest itself in an analysis of
optical images, which agrees with the general lack of excess found by
the many authors who relied upon optical imaging for their bar
classification (see references above). Laurikainen et al. (2004) also
find that the bars in active galaxies are weaker than those in
non-active galaxies, a result which confirms earlier indications to
this effect by Shlosman, Peletier \& Knapen (2000) and by Laurikainen,
Salo, \& Rautiainen (2002).

We can thus conclude that there is a slight, though significant,
excess of bars among Seyfert galaxies as compared to non-active
galaxies. This result is found only when using near-IR images, and
only when applying rigorous and objective bar classification
methods. Even so, there remain important numbers of active galaxies
without any evidence for a bar, and, on the other hand, many
non-active galaxies which do have apparently suitable bars. Given that
any fuelling process must be accompanied by angular momentum loss,
most easily achieved by gravitational non-axisymmetries, either the
timescales of bars (or interactions, see below) are different from
those of the activity, or the non-axisymmetries are not as easy to
measure as we think, for instance because they occur at spatially
unresolvable scales, and could be masqueraded to a significant extent
by, e.g., dust or star formation (Laine et al. 2002), or because they
occur in the form of weak ovals (e.g., Kormendy 1979) which will not
necessarily be picked up by ellipse fitting or Fourier
techniques. Additional work is clearly needed, but it is not clear
whether this should be aimed primarily at the large-scale bars
described in this Section, or perhaps better at the kinematics and
dynamics of the very central regions of active and non-active
galaxies. In any case, the use of carefully matched samples and
control samples is of paramount importance.

\section{The effects of interactions}

Galaxy interactions can easily lead to non-axisymmetries in the
gravitational potential of one or more of the galaxies involved, and
as such can be implicated in angular momentum loss of inflowing
material, and thus conceivably in starburst and AGN fuelling (e.g.,
Shlosman et al. 1989, 1990; Mihos \& Hernquist 1995).

\subsection{Interactions and starburst activity}

It is well known that there is ample anecdotal evidence for the
connection between galaxy interactions and starburst activity. This is
perhaps clearest for the most extreme infrared sources, specifically
the Ultra-Luminous InfraRed Galaxies (ULIRGs). They are powered mainly
by starbursts (Genzel et al. 1998), and it has been known since
briefly after their discovery that they occur in galaxies with
disturbed morphologies, presumably as a result of recent interactions
(e.g., Joseph \& Wright 1985; Armus, Heckman, \& Miley 1987; Sanders
et al. 1988; Clements et al. 1996; Murphy et al. 1996; Sanders \&
Mirabel 1996).  Given that the ULIRGs are both among the most extreme
starbursts known, and are occurring in interacting galaxies, one can
infer that such massive starbursts are in fact powered by gas which
has lost angular momentum in galaxies which are undergoing a major
upheaval, i.e., are merging or interacting.

More in general, and considering galaxies less extreme than those in
the ULIRG class, there is considerable evidence for a connection
between interactions and enhanced star formation in galaxies, often
measured using galaxy colours which are bluer in the case of current
star formation (see, e.g., the seminal paper by Larson \& Tinsley
1978). But even so, a more detailed consideration can expose possible
caveats. We mention the recent paper by Bergvall, Laurikainen, \&
Aalto (2003), who considered two matched samples of nearby interacting
(pairs and clear cases of mergers) and non-interacting galaxies, and
measured star formation indices based on $UBV$ colours. From this
analysis, Bergvall et al. do {\it not} find evidence for significantly
enhanced star-forming activity among the interacting/merging galaxies,
although they do report a moderate increase in star formation in the
very centres of their interacting galaxies. Interesting in this
respect are also recent results from a combination of Sloan Digital
Sky Survey and 2dF Galaxy Redshift Survey data, presented by Balogh et
al. (2004). These authors study the equivalent width of H$\alpha$
emission, a measure of starburst activity, and find no correlation
between its distribution among the star-forming population of galaxies
and the environment.

So although mergers can undoubtedly lead to massive starbursts, they
appear to do so only in exceptionally rare cases. Bergvall et
al. (2003) estimate that only about 0.1\% of a magnitude limited
sample of galaxies will host massive starbursts generated by
interactions and mergers. Most interactions between galaxies may not
lead to any increase in the starburst activity. Those that do may be
selected cases where a set of parameters, both internal to the
galaxies and regarding the orbital geometry of the merger, is
conducive to the occurrence of starburst activity (e.g., Mihos \&
Hernquist 1996). To further illustrate this point, we quote the
results published by Laine et al. (2003), who find very little
evidence for trends in starburst activity from detailed {\it HST}
imaging of the Toomre sequence of merging galaxies.

\subsection{Interactions and Seyfert activity}

Interactions and mergers have long been suspected of triggering
high-lumi\-no\-si\-ty AGN such as QSOs (e.g., Disney et al. 1995;
Bahcall et al. 1997), although many of such AGN seem to lie in
entirely undisturbed elliptical systems. In fact, Dunlop et al. (2003)
show that the host galaxy properties of radio-loud and radio-quiet AGN
are indistinguishable from those of quiescent but otherwise comparable
galaxies, and Floyd et al. (2004) find no correlation between the
luminosity of a quasar and the presence of any morphological
disturbance in the host.

Seyfert activity is known to occur in interacting and merging
galaxies, and several rather spectacular examples are well known (for
instance NGC~2992, or a number of the closest ULIRGs such as Mrk~273).
To check statistically whether there is a connection between
interactions and the occurrence of this type of nuclear activity,
authors have considered the numbers of companions to Seyfert galaxies
as compared to non-active control galaxies (e.g., Fuentes-Williams \&
Stocke 1988; de Robertis, Yee, \& Hayhoe 1998; Schmitt 2001), or,
alternatively, have searched for different fractions of Seyfert or AGN
activity among more or less crowded environments (e.g., Kelm, Focardi,
\& Palumbo 1998; Miller et al. 2003). The conclusion from this
substantial body of work must be that no unambiguous evidence exists
for a direct connection between the occurrence of Seyfert activity and
interactions.  Some earlier work did report claims of a statistical
connection, but this work was unfortunately plagued by poor control
sample selection (see Laurikainen \& Salo 1995 for a detailed review),
and most early studies, but also some of the recent ones, are not
based on complete sets of redshift information for the possible
companion galaxies.  In addition, Laine et al. (2002) have shown that
the bar fraction among both the Seyfert and non-Seyfert galaxies in
their sample is completely independent of the presence of companions
(interacting galaxies were not considered by Laine et al.).  We thus
conclude that interactions and Seyfert activity may well be linked in
individual cases, but that as yet the case that they are statistically
connected has not been made convincingly.


\section{Bars and nuclear rings}

Apart from concentrating gas in the central regions of galaxies, as
discussed in Section~2, bars also set up resonances which can act as
focal points for the gas flow. As reviewed by, e.g., Shlosman (1999),
gas concentrates there in limited radial ranges, where it can become
gravitationally unstable and form stars. Rings in disk galaxies are
mostly identified by their star formation, either by their blue
colours or by H$\alpha$ emission, and are intimately linked to the
internal dynamics and the evolution of their hosts (see Buta \& Combes
1996 for a comprehensive review on galactic rings).

\begin{figure}[ht]
\vskip.2in
\centerline{\includegraphics[width=4in]{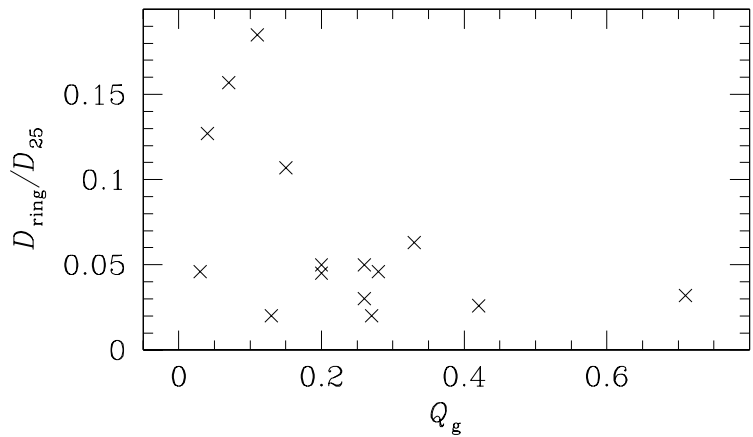}}
\caption{Relative size (ring diameter divided by host galaxy diameter)
for a sample of 15 nuclear rings as a function of the gravitational
torque $Q_{\rm g}$, or strength, of the bar of its host galaxy. Data
from Knapen, P{\' e}rez-Ram{\'{\i}}rez, \& Laine (2002) and Knapen
(2004b).}
\label{ringdiam}
\end{figure}

Nuclear rings are those on scales of less than one to roughly two
kiloparsec in radius. They are rather common, and occur in about 20\%
of nearby spiral galaxies (Knapen 2004b).  They can be directly linked
to inner Lindblad resonances (Knapen et al. 1995a; Heller \& Shlosman
1996; Shlosman 1999), and are in fact found almost exclusively in
barred galaxies (e.g., Buta \& Combes 1996; Knapen 2004b; but see
below for a counterexample). Individual gas clouds in a nuclear ring
can undergo a Jeans-type collapse either spontaneously (Elmegreen
1994), or after compression by density waves set up by the bar (Knapen
et al. 1995a; Ryder, Knapen, \& Takamiya 2001).  Nuclear rings are
thus not only excellent tracers of massive star formation in starburst
regions, but also of the dynamics of their host galaxies. The latter
point is illustrated by Fig.~\ref{ringdiam}, which shows that nuclear
rings with a large size relative to their host galaxy can only occur
in bars with a low gravitational torque, or: {\it large rings cannot
occur in strong bars}. This confirms the results from earlier theory
and modelling that the extent of the perpendicular $x$2 orbits needed
to sustain the nuclear ring is limited as the bar gets stronger, i.e.,
as the $x$1 orbits become more elongated (see Knapen et al. 1995a;
Heller \& Shlosman 1996; Knapen, P\'erez-Ram\'\i rez, \& Laine 2002;
Knapen 2004b), and graphically shows how intricately bars and nuclear
rings are related.

A small number of rings, or pseudo-rings, apparently occur in
non-barred galaxies. Some of these hosts, although classified as
non-barred from optical imaging in the major catalogues, are obviously
barred when imaged in the near-IR (e.g., NGC~1068, Scoville et
al. 1988, and NGC~4725, Shaw et al.  1993; M\"ollenhoff, Matthias, \&
Gerhard 1995). In other cases, a non-barred host galaxy may either
have an oval distortion, or be undergoing the effects of an
interaction with a companion galaxy. In either of these cases, the
gravitational potential of the galaxy could be disturbed, and the
non-axisymmetric potential could lead to ring formation, in much the
same way as in the presence of a bar potential (Shlosman et al. 1989).

\begin{figure}[ht]
\vskip.2in
\caption{Real-colour image of the galaxy NGC~278 produced from
archival $F450W$, $F606W$, and $F814W$ $HST$ images.  Area shown is
about 100~arcsec on a side, or about 5.7~kpc. The blue region,
apparently made up of spiral arm fragments, indicates the nuclear ring,
with a radius of 1.1~kpc. Image data from Knapen et al. (2004a).}
\label{278fig}
\end{figure}

A nice example is that of NGC~278, a small, nearby and isolated spiral
galaxy ($v_{\rm sys}=640$\,km\,s$^{-1}$; $D=11.8$~Mpc;
$D_{25}=7.2$~kpc; $M_B=-18.8$).  Although classified as SAB(rs)b in
the RC3, there is no evidence for the presence of a bar in this galaxy
from either {\it HST} WFPC2 or ground-based NIR imaging (Knapen et
al. 2004a).  The optical disk of NGC~278 shows two distinct regions,
an inner one with copious star formation and clear spiral arm
structure, shown in Fig.~\ref{278fig}, and an outer one ($r>27$~arcsec
or $r>1.5$~kpc) which is almost completely featureless, of low surface
brightness, and rather red. NGC~278 has a large \hi\ disk, which is
morphologically and kinematically disturbed, as seen from \hi\ data
(Knapen et al. 2004a). These disturbances suggest a recent minor
merger with a small gas-rich galaxy, perhaps similar to a Magellanic
cloud.

The scale and morphology of the region of star formation in NGC~278
indicate that this is in fact a nuclear ring, albeit one with a much
larger {\it relative} size with respect to its host galaxy than
practically all other known nuclear rings (the absolute radius of the
nuclear ring is about a kiloparsec, normal for nuclear rings). Knapen
et al. (2004a) postulate that it is in fact the past interaction which
has set up a non-axisymmetry in the gravitational potential, which in
turn, in a way very similar to the action of a classic bar, leads to
the formation of the nuclear ring. The case of NGC~278 illustrates how
in apparently non-barred galaxies rings can be caused by departures
from axisymmetry induced by interactions, but also shows how difficult
it can be to uncover this: in the case of NGC~278 only through
detailed \hi\ observations.

\section{Rings and nuclear activity}

Although nuclear rings and especially nuclear activity have as
separate topics received considerable attention in the literature,
their possible interrelation has not been much studied. Many nuclear
rings, of course, are anecdotally known to occur in galaxies which
also host a nuclear starburst or a prominent AGN (often of Seyfert or
LINER type, given the typical parameters of the host galaxies
involved), and some famous examples include NGC~1068 and NGC~4303.

In a recent paper, we explored the correlations between nuclear
activity (both of the non-stellar and starburst variety) and the
occurrence of nuclear rings in a sample of 57 nearby spiral galaxies
(Knapen 2004b). Using information on the activity from the NASA/IPAC
Extragalactic Database (NED) and ring parameters as derived from new
H$\alpha$ imaging (Knapen et al. 2004b), we found not only that
nuclear rings significantly more often than not occur in galaxies
which also host nuclear activity (only two of the 12 nuclear rings
occur in a galaxy which is neither a starburst nor an AGN host; 30 of
the 57 sample galaxies overall would fall into this category), but
also that the circumnuclear H$\alpha$ emission morphology of the AGN
and starbursts is significantly more often in the form of a ring than
in non-AGN, non-starburst galaxies (38\% of AGN, 33\% of starbursts,
11\% of non-AGN, and 7\% of non-AGN non-starburst galaxies have
circumnuclear rings in our sample of galaxies).

Although the number of galaxies in this initial study is rather small
for detailed statistical analyses, we did find this most interesting
correlation between the occurrence of nuclear rings and that of
nuclear activity. Our initial interpretation of this effect is that
both nuclear rings, as traced by the massive star formation within
them, and starbursts and AGN (of the Seyfert or LINER variety) trace
very recent gas inflow.  Although it is not clear {\it a priori} why
the kpc-scale fuelling of nuclear rings and the pc-scale fuelling of
activity might be so closely related, nuclear rings do seem to show a
potential as tracers of AGN fuelling. These findings may also be
related to the reported higher incidence of rings among Seyfert and
LINER hosts than among non-AGN galaxies (Arsenault 1989 for nuclear
rings; Hunt \& Malkan 1999 for inner and outer rings). All these
aspects of rings and nuclear activity need further scrutiny.

\section{Concluding remarks}

From theory and modelling, and increasingly also from observations, it
is clear that bars can remove angular momentum from gaseous material,
and thus drive it from the disk into the central kpc-scale regions of
a galaxy (Section~2). In contrast, the evidence that this centrally
condensed gas directly and immediately leads to AGN or starburst
activity remains rather elusive.  The relevant results, reviewed more
in depth elsewhere in this paper as indicated below, have been
summarised in Table~1, where bars and interactions have been labelled
as primary indicators for links between the inflow-provoking
mechanisms and the possibly resulting AGN or starburst activity.  Also
listed in Table~1 is a small number of so-called secondary indicators,
which have received attention as outlining possible links between
inflow and activity, but which may well be a result of one of the
primary indicators. The information summarised in Table~1 can be
related to the content of this paper as follows:

\begin{table}[ht]
\caption[]{Summary of observational evidence for relations between
various host galaxy features and Seyfert/LINER and starburst activity}
\begin{tabular*}{\textwidth}{@{\extracolsep{\fill}}lcc}
\sphline
\it Feature&\it Seyferts/LINERs &\it Starbursts\cr
\sphline
\multicolumn{3}{c}{\underline{Primary indicators}}\\
&&\\
Bars &         yes (but 2.5$\sigma$) & yes (but not in general?)\\
Interactions & no                    & yes (but extremes only?)\\
&&\\
\multicolumn{3}{c}{\underline{Secondary indicators}}\\
&&\\
Nuclear bars & no                    & N/A\\
Rings        & yes, some             & yes (nuclear rings at least)\\
\sphline
\end{tabular*}
\end{table}

\begin{itemize}

\item Starbursts can be provoked by bars or interactions, at least in
some cases (Section~2,1, 3.1).

\item There is an increasing body of evidence that Seyfert activity
preferentially occurs in barred host galaxies, but the effect is
a statistical one, and not very pronounced (Section~2.2).

\item There is no convincing evidence that AGN hosts are
interacting more often than non-AGN (Section~3.2). 

\item Nuclear bars (Section~2.2) have great theoretical potential for
bringing fuel very close to the centre of a galaxy (the ``bars within
bars'' scenario, Shlosman et al. 1989) but have so far not lived up to
that potential in terms of their detections in imaging surveys, where
no higher nuclear bar fractions have been found in Seyferts as
compared to non-Seyferts. As far as we aware, the possible statistical
connections between nuclear bars and starbursts have not yet been
studied.

\item There is some interesting evidence that rings, both
nuclear and non-nuclear, may be related to the presence of
low-luminosity AGN activity (Section~5). This issue needs further
exploration, but in any case the rings will most likely have formed
under the influence of either a bar or another form of non-axisymmetry
in the gravitational potential of the host, hence the inclusion of
rings as secondary indicators in Table~1. 

\item Finally, there is a direct link between starburst activity and
the presence of nuclear rings, since small nuclear rings with
significant massive star formation can be classified as starburst, and
since many starbursts might in fact be circumnuclear rather than
nuclear, albeit with small ring radii of tens to hundreds of parsec
(e.g., Gonz\'alez-Delgado et al. 1998), or possibly even
smaller. Statistical links between inner/outer rings and starburst
activity have not yet been explored.

\end{itemize}

Since we have known for quite some time that net radial gas inflow
must be accompanied by the loss of significant quantities of angular
momentum, and that the kind of deviations from axisymmetry in the
gravitational potential of the host galaxy set up by bars and
interactions is well suited to lead to such angular momentum loss (see
reviews by Shlosman et al. 1989, 1990; Shlosman 2003), we must be
missing some part of the puzzle. One possibility is that we are not
looking at the right things at the right time: the spatial- as well as
the time-scales under consideration may not be correct. So far, the
spatial scale considered observationally has been from tens of kpc
down to, roughly, a few hundred parsec. Whereas this range may be
wholly adequate for the study of a major starburst, which can span up
to a kpc, it may well be wholly {\it in}adequate for AGN fuelling,
which is expected to be related to accretion to a SMBH, on scales of
AUs. As far as timescales are concerned, what has been considered in
the studies reviewed here is generally a rather long-lived phenomenon
influencing kpc-scale regions (bar or galaxy-galaxy
interaction). Starburst or AGN activity, on the other hand, occurs on
essentially unknown timescales (somewhere around $10^6-10^8$ years
could be expected for most AGN or starbursts). If the starburst or AGN
activity is indeed short-lived, and possibly also periodic, the
connection between the presence of activity at the currently observed
epoch and any parameter of the host galaxy is not necessarily
straightforward (as pointed out, e.g., by Beckman 2001).

The fact that we see any correlations at all, such as those of bar
fractions with the presence of starburst or Seyfert activity,
indicates that bars and interactions do have a role, presumably by
establishing a gas reservoir in the central kpc region. In the coming
years, we must start to disentangle the effects of gravitationally
induced gas inflow, which brings gas to the inner kpc region at least,
from those of possible other mechanisms which can transport the gas
further in, and from the time scales and duty cycles of the
activity. We seem to have reached the limits of purely morphological
studies of the central regions of active and non-active galaxies
(e.g., Laine et al. 2002), and must start to worry about the effects
of the AGN or starburst on their immediate surroundings as we push the
observations to smaller spatial scales, of tens of parsecs. One must, hence, 
move on to careful studies of the gas and stellar kinematics and dynamics.
Integral field spectroscopy (e.g., Bacon et al. 2001), especially when
used in conjunction with adaptive optics techniques, should allow a
good deal of progress here, giving simultaneous high-resolution
mapping of the distributions of stellar populations and dust, as well
as of the gas and stellar kinematics. In combination with detailed
numerical modelling, this could lead to the detection of the dynamical
effects of, e.g, nuclear bars on gas flows which may be more directly
related to the fuelling process of starbursts and/or AGN.

{\it Acknowledgements} I am indebted to my collaborators on the
various aspects of the work described here, especially John Beckman,
Shardha Jogee, Seppo Laine, Reynier Peletier, and Isaac
Shlosman. Valuable comments by conference participants have helped
improve this paper.

\end{document}